\documentclass{aastex}
\usepackage{emulateapj5}

\newcommand{\approxlt}{\mbox{$\;^{<}\hspace{-0.24cm}_{\sim}\;$}}
\newcommand{\approxgt}{\mbox{$\;^{>}\hspace{-0.24cm}_{\sim}\;$}}

\slugcomment{Accepted by the Astrophysical Journal}

\shorttitle{A Comparison of Crab and Vela Glitch Parameters with
Neutron Star Models}
\shortauthors{Crawford \& Demia\'nski}

\begin{document}

\title{A Comparison of Measured Crab and Vela Glitch Healing
Parameters with Predictions of Neutron Star Models}

\author{Fronefield Crawford}
\affil{Department of Physics, Haverford College, Haverford, PA 19041, USA}
\email{fcrawfor@haverford.edu} 

\author{Marek Demia\'nski}
\affil{Institute of Theoretical Physics, University of Warsaw, Warsaw, Poland}
\affil{Department of Astronomy, Williams College, Williamstown, MA 01267, USA}
\email{Marek.Demianski@fuw.edu.pl} 

\begin{abstract}
There are currently two well-accepted models that explain how pulsars
exhibit glitches, sudden changes in their regular rotational
spin-down. According to the starquake model, the glitch healing
parameter, $Q$, which is measurable in some cases from pulsar timing,
should be equal to the ratio of the moment of inertia of the
superfluid core of a neutron star (NS) to its total moment of
inertia. Measured values of the healing parameter from pulsar glitches
can therefore be used in combination with realistic NS structure
models as one test of the feasibility of the starquake model as a
glitch mechanism. We have constructed NS models using seven
representative equations of state of superdense matter to test whether
starquakes can account for glitches observed in the Crab and Vela
pulsars, for which the most extensive and accurate glitch data are
available.  We also present a compilation of all measured values of
$Q$ for Crab and Vela glitches to date which have been separately
published in the literature. We have computed the fractional core
moment of inertia for stellar models covering a range of NS masses and
find that for stable NSs in the realistic mass range $1.4 \pm
0.2~M_{\odot}$, the fraction is greater than 0.55 in all cases. This
range is not consistent with the observational restriction $Q
\approxlt 0.2$ for Vela if starquakes are the cause of its
glitches. This confirms results of previous studies of the Vela pulsar
which have suggested that starquakes are not a feasible mechanism for
Vela glitches. The much larger values of $Q$ observed for Crab
glitches ($Q \approxgt 0.7$) are consistent with the starquake model
predictions and support previous conclusions that starquakes can be
the cause of Crab glitches.
\end{abstract}


\section{Introduction} 

Soon after the discovery of radio pulsars \citep{hbp+68}, they were
identified as rotating, highly magnetized neutron stars (NSs)
\citep{g68, g69, og69}. It was observationally established that pulsar
periods slowly and monotonically increase with time as a result of
magnetic braking and rotational energy loss.  Early in 1969, it was
noticed that the Vela pulsar suddenly increased its angular velocity
\citep{rd69, rm69}.  These distinct and sudden increases in rotational
frequency, known as glitches, were later regularly observed in the
Vela and Crab pulsars and, more infrequently, in other pulsars (e.g.,
Shemar \& Lyne 1996\nocite{sl96}; Lyne, Shemar, \& Smith
2000\nocite{lss00}). The Crab (PSR B0531+21) and Vela (PSR B0833$-$45)
pulsars are the two best-studied glitching pulsars since they have
each been observed to glitch a number of times and are bright and easy
to monitor. These pulsars have the most extensive and accurate glitch
data published and currently provide the best test of the physical
mechanisms by which pulsar glitches occur.

There are currently two competing models that are well-accepted to
explain how pulsars glitch.  In simplest terms, both models treat the
NS as a two-component body, with a superfluid interior core surrounded
by a rigid external crust (e.g., Ruderman 1972\nocite{r72}). These two
components are dynamically weakly coupled through the magnetic field.

In the starquake glitch model \citep{r76, bp71,acc+96}, a slight
equatorial oblateness in the crust can be formed if the NS is born
rapidly spinning. As the pulsar slows down via magnetic braking, the
deformation cannot be supported through centrifugal pressure from the
core, and the crust can suddenly crack under gravity.  The subsequent
reduction in oblateness reduces the moment of inertia, resulting in a
sudden increase in the rotational frequency, seen as a glitch.

Alternatively, in the vortex unpinning model of glitches \citep{ai75,
aap+84, acc+93}, angular momentum is stored in vortices of superfluid
which migrate outward from the core as the star slows down. These
vortices can become pinned to nuclei in the transition region between
the core and outer crust, thereby preventing angular momentum transfer
to the crust. A differential rotation develops between the core and
crust until a catastrophic unpinning of the vortices occurs. Angular
momentum is suddenly transferred to the crust, spinning it up. Since
the crust is tightly coupled to the external magnetic field, this
sudden spin-up in the rotation period is observed as a glitch.

Within the starquake model, the time evolution of a pulsar's
rotational frequency after the occurrence of a glitch \citep{wmp+00}
can be fit by the equation:

\begin{equation} 
\Omega(t) = \Omega_{0}(t) + \Delta \Omega_{t=0} \left[ 1 - Q
(1-e^{-t/\tau}) \right] + \Delta \dot{\Omega}_{p} t .
\label{eqn-glitch}
\end{equation} 

Here $\Omega_{0}(t)$ represents an extrapolation of the pre-glitch
frequency evolution, $\Delta \Omega_{t=0}$ is the magnitude of the
change in the rotational frequency at the time of the glitch ($t=0$),
$\tau$ is a characteristic exponential healing or recovery time after
the glitch, and $Q$ is the fraction of the initial frequency change
that is eventually recovered (known as the healing parameter). The
last term accounts for any permanent (not recovered) change $\Delta
\dot{\Omega}_{p}$ in the frequency derivative after the glitch.  The
parameter $Q$ in Equation \ref{eqn-glitch} can be determined
observationally \citep{st83} from the jump in $\Omega$ and its first
and second time derivatives:

\begin{equation}
Q = \frac{\Delta \dot{\Omega}_{t=0}^{2}}{\Delta \ddot{\Omega}_{t=0}
\Delta \Omega}_{t=0} .
\end{equation} 

In the starquake model, $Q$ is related to the moments of inertia of
the components of the star according to \citep{psr74}:

\begin{equation}
\label{eqn-Q}
Q = \frac{I_{\rm core}}{I_{\rm total}} ,
\end{equation}

where $I_{\rm core}$ and $I_{\rm total}$ are respectively the moment
of inertia of the superfluid core and the moment of inertia of the
entire star. We have constructed a series of models of the NS interior
using representative high-density equations of state (EOSs) in order
to test whether the calculated moments of inertia in the models
satisfy the prediction of the starquake model for Crab and Vela
glitches.  We have compared the model results with the measured values
of $Q$ from Crab and Vela glitches to see if they are consistent.

\section{Equations of State}

We have constructed NS models using seven representative EOSs of
superdense matter from which parameterized model stars could be
produced. For densities $\rho \approxlt 1.6 \times 10^{14}$ g
cm$^{-3}$ (crustal densities), we have used three EOSs for three
separate density regimes (see, e.g., Shapiro \& Teukolsky
1983\nocite{st83}). These three EOSs are described by \citet{fmt49},
\citet{bps71}, and \citet{bbp71}. The seven high-density EOSs, which
dominate the macroscopic characteristics of the model stars, are
briefly described below. Further details about each EOS can be found
in the sources referenced.

\subsection{The BJW Equation of State}  

For densities above the nuclear density, a combination of EOSs
presented by \citet{c75} was used for the modeling. For densities up
to $\sim 5 \times 10^{15}$ g cm$^{-3}$, the EOS described by
\citet{bj74} was employed. This EOS is an improvement upon the work of
\citet{r68} which includes a repulsive nucleonic core arising from the
exchange of vector mesons in a hyperonic liquid. At higher densities,
the EOS of \citet{w74} was used, although the NS models produced in
this density regime were beyond the stability limit. In this EOS,
nucleons interact attractively via exchange of scalar mesons and
repulsively via exchange of more massive vector mesons. Summaries of
these EOSs are also presented by \citet{st83}. We call this EOS BJW.

\subsection{The FPS Equation of State} 

\citet{lrp93} refer to an EOS calculated by \citet{fp81} for
high-density neutron and nuclear matter. This EOS employs the
microscopic V14 two-body potential, a three nucleon interaction (TNI)
potential, and uses hypernetted chain techniques. This EOS was
modified by fitting the microscopic interaction of Friedman \&
Pandharipande to a Skyrme-like energy density function
\citep{s59}. The essential feature of the Skyrme model is a two-body
interaction that has the spatial character of a two-body delta
function plus derivatives. We refer to this EOS as FPS.

\subsection{GWM Equation of State} 

This EOS is based on a variant of the theory of nuclear field coupling
\citep{zm90} in which the scalar field is coupled to the derivative of
the nucleon field. A modification proposed by \citet{gwm92}, called
the hybrid derivative coupling model, replaces the purely derivative
coupling of the scalar field to baryons and vector mesons with a
Yukawa point and derivative coupling to baryons and both vector
fields. The coupling model is consistent with the experimentally
inferred binding energy of lambda hyperons in nuclear matter. The
resulting EOS, called GWM here, maintains equilibrium between all
baryons to convergence and leptons. The existence of hyperons
significantly softens this EOS.

\subsection{HKP Equation of State}

This EOS, proposed by \citet{hkp81}, supposes that a minimal condition
for the properties of high-density matter is that it must produce the
observed values for the saturation properties of nuclear matter. The
relativistic mean field theory of \citet{s79} is assumed in which
nucleon interactions are governed by the exchange of neutral scalar
and neutral vector mesons, pions, and rho mesons. This theory is an
extension of the relativistic mean field theory of \citet{w74}. The
authors assume a conservative range of possible nuclear saturation
densities, the mean of which is $2.8\times 10^{14}$ g cm$^{-3}$. The
EOS used here, called HKP, is based on this nuclear saturation value.

\subsection{WFF Equation of State}

\citet{wff88} describe a microscopic EOS of dense nuclear matter
constrained by nucleon-nucleon scattering data. The interaction
includes a two-nucleon Urbana $v_{14}$ (UV14) potential and a TNI
three-body potential term \citep{lp81a, lp81b}.  The three-body
potential includes a repulsive term, the primary effect of which is a
reduction in the intermediate-range attraction of the two-nucleon
potential, and an attractive term, which becomes negligible at high
densities. We refer to this EOS as WFF.

\subsection{APR Equations of State} 

\citet{apr98} describe a set of realistic EOSs based on the Argonne
$v_{18}$ (A18) two-nucleon interaction \citep{wss95}, calculated using
variational chain summation methods. This is supplemented with a
relativistic boost correction term \citep{fpf95} and a TNI term based
on the Urbana IX model \citep{ppc+95}. The two cases of pure neutron
matter (PNM) and symmetric nuclear matter (SNM) are separately
considered. The latter is composed of equal numbers of neutrons and
protons in beta equilibrium.  The two EOSs used to construct the
models presented here were derived from a parabolic fit to and
extrapolation of the tabulated nucleon density and energy values given
for PNM and SNM by \citet{apr98}.  We refer to the two EOSs based on
PNM and SNM as APR(p) and APR(s), respectively.

\section{The Modeling Procedure} 

To construct parameterized NS models for each EOS, we first chose a
central density $\rho_{c} \equiv \rho(0)$ for each stellar model. The
EOS and the relativistic expressions for stellar structure
\citep{ov39} were used to calculate the pressure gradient and
incremental mass contained in concentric shells as the model iterated
outward from the center of the star. The relativistic structure
equations are:

\begin{equation} 
\frac{dm(r)}{dr} = 4\pi r^{2} \rho(r) \\ \nonumber \\
\end{equation} 

\begin{equation} 
\frac{dP(r)}{dr} = -\left(\rho(r) + \frac{P(r)}{c^{2}}\right)
\frac{G\left[m(r) + 4\pi r^{3} P(r)/c^{2}\right]}
{r^{2}\left[1-2Gm(r)/rc^{2}\right]} ,
\end{equation}

\noindent
where $m(r)$ represents the mass of the star internal to a distance
$r$ from the center of the star. When the pressure reached zero, $P(R)
= 0$, the edge of the star was reached ($r = R$) and the relevant
macroscopic parameters could be read off.

To calculate the relativistic moment of inertia $I(r)$ interior to a
radius $r$, expressions given by \citet{ab77} were used which account
for the Lense-Thirring frame-dragging effect (e.g., Glendenning
2000\nocite{g00}) (see discussion below):

\begin{equation} 
\frac{dI(r)}{dr} = \frac{8\pi}{3} \frac{[\rho(r) + P(r)/c^{2}] r^{4}
e^{-\phi(r)}}{[1 - 2Gm(r)/rc^{2}]^{1/2}} 
\frac{\bar{\omega}(r)}{\Omega} .
\label{eqn-moment}
\end{equation} 

Here $\Omega$ is the observed stellar rotational frequency and
$\bar{\omega}(r)$ is the angular velocity of the star at a distance
$r$ relative to the angular velocity of the rotating local inertial
frame, which is dragged at angular velocity $\omega(r)$:

\begin{equation}
\bar{\omega}(r) = \Omega - \omega(r) . 
\end{equation}

The term $e^{-\phi(r)}$ in Equation \ref{eqn-moment} translates
$\bar{\omega}(r)$ as measured from infinity to $\bar{\omega}(r)$ as
measured in the local inertial frame at $r$. The following auxiliary
relations were used to determine values of $\phi(r)$ and
$\bar{\omega}(r)$ throughout the star ($r < R$) in the moment of
inertia calculation:

\begin{equation} 
\frac{d\phi(r)}{dr} = - \left( \frac{1}{\rho(r)c^2 + P(r)} \right)
\frac{dP(r)}{dr}
\end{equation} 

\begin{equation} 
j(r) = \left( 1 - \frac{2Gm(r)}{rc^{2}} \right) e^{-\phi(r)} 
\end{equation} 

\begin{equation} 
\frac{d}{dr} \left( r^{4} j(r) \frac{d \bar{\omega}(r)}{dr} \right) + 4 r^{3} \frac{dj(r)}{dr} \bar{\omega}(r) = 0 .
\end{equation} 

An arbitrary value of $\bar{\omega}(0)$ was chosen and the following
boundary conditions imposed in the models:

\begin{equation} 
\frac{d \bar{\omega}(0)}{dr} = 0 
\end{equation} 

\begin{equation} 
\bar{\omega}(R) = \Omega - \frac{R}{3} \frac{d \bar{\omega}(R)}{dr} . 
\end{equation} 

This latter condition (in which a value for $\Omega$ is determined at
the end of the model iteration) was used to scale the arbitrarily
chosen $\bar{\omega}(0)$ in such a way as to yield the proper observed
$\Omega$ in a subsequent modeling pass. This subsequent pass produced
the correct $\bar{\omega}(r)$ for all $r < R$. The condition of
finite $\phi(0)$ was also imposed.

Milne's centered algorithm \citep{hww+65} was used for the
iteration. Since there is a transition region between the superfluid
core and the rigid crust which consists of a mixture of nuclei and
free baryons, there is no clear division where the core-crust
transition occurs. We chose a transition density near the nuclear
density ($\rho_{\rm transition} \sim 2.4 \times 10^{14}$ g cm$^{-3}$)
in order to separate the two components in our models. From this we
obtained $I_{\rm core} \equiv I(r_{\rm transition})$ and $I_{\rm
total} \equiv I(R)$ for each model NS and could compare the ratio of
these to the measured values of $Q$ from glitch observations.  In
previous work, some authors (e.g., Datta \& Alpar 1993\nocite{da93};
Shapiro \& Teukolsky 1983\nocite{st83}) have assumed a slightly lower
core-crust transition density than the one used here. This lower
transition density reduces the size of the crust, thereby increasing
the fractional moment of inertia of the core. Although slight
differences in the choice of the transition density do not affect our
model results, a significant reduction would lead to a more stringent
test of the starquake model (i.e., it would further limit the range of
observed $Q$ that would be consistent with the predictions of the
starquake model). Thus, our choice of $\rho_{\rm transition}$ is a
conservative one for this purpose.

For angular rotation speeds comparable to those of the Crab and Vela
pulsars ($\Omega \sim 190$ rad s$^{-1}$ and $70$ rad s$^{-1}$,
respectively), the macroscopic star parameters are affected by
first-order rotation effects, but not significantly so by second-order
effects.

The Lense-Thirring frame dragging effect is a first-order effect which
scales with rotational speed as $(\Omega / \Omega_{c})$ (Arnett \&
Bowers 1977), where $\Omega_{c} = (GM/R^{3})^{1/2}$, the critical
value above which centrifugal mass shedding would occur for the
rotating star. For realistic NS models, $\Omega_{c} \sim 13000$ rad
s$^{-1}$. This first-order effect is therefore $\sim$ 1-2\% for the
Crab and Vela pulsars and is taken into consideration in the
calculation of the relativistic moment of inertia in the models.

NS mass shifts and deformations in sphericity from rotation (which
would complicate the simple spherical stellar modeling presented here)
are second order effects, scaling as $(\Omega / \Omega_{c})^{2}$
(Arnett \& Bowers 1977). This term would be significant for
millisecond pulsars, but for the Crab pulsar, the second-order effect
is $\sim 10^{-4}$, and it is even lower for Vela. Thus, the
second-order effect can be safely ignored when considering the
macroscopic parameters of our NS models.

\section{Results} 

We have produced a set of parameterized NS models using the seven
high-density EOSs described above. In order to decide which models in
each series were appropriate to use in the comparison with the
observed values of $Q$, we used a realistic NS mass range as a
reasonable constraint.  Models with a NS mass outside of the range
were not used in the comparison.

\citet{tc99} have presented a statistical study of NS masses for the
known binary radio pulsar population for which useful mass constraints
could be derived. They find that, statistically, the distribution of
pulsar masses falls in the range $M = 1.35 \pm 0.04~M_{\odot}$. This
is consistent with the initial NS mass function from simulations of
supernovae, which predict that NSs are formed with $M \approxgt
1.2~M_{\odot}$ \citep{tww96}. For our comparison and analysis, we
consider only the mass range $M = 1.4 \pm 0.2~M_{\odot}$, which we
assume to be a conservative estimate of the realistic NS mass range.

\begin{figure}
\epsscale{0.6} 
\plotone{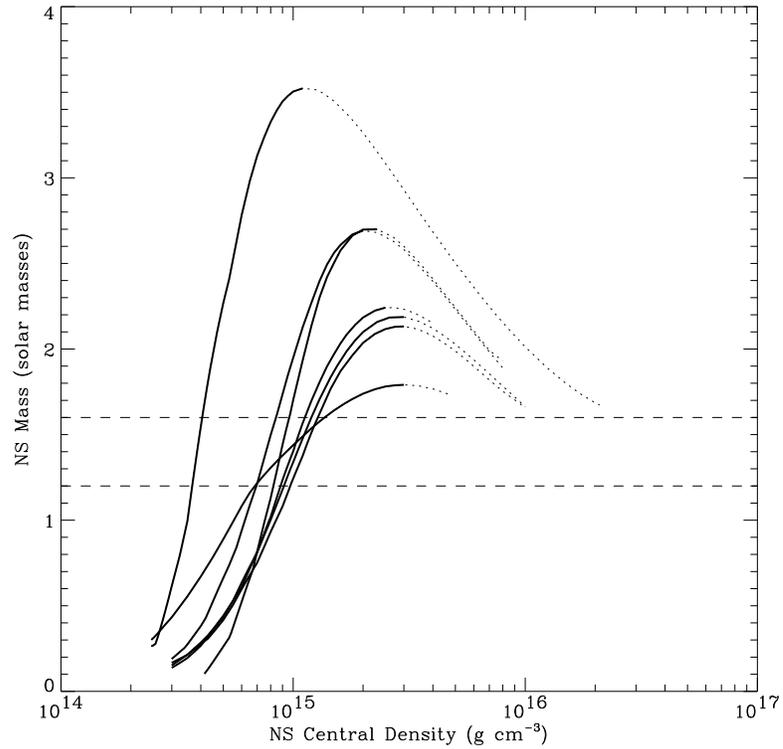} \figcaption[f1.eps]{Mass vs.~central density for NS
models constructed from seven high-density EOSs. The seven parameter
curves correspond to these seven EOSs in order of increasing maximum
mass: GWM, BJW, FPS, WFF, APR(p), APR(s), and HKP.  Branches with
positive slope (a necessary condition for stability) are indicated
with bold lines. The dashed horizontal lines correspond to the mass
range $1.4 \pm 0.2~M_{\odot}$, which is assumed to be a realistic NS
mass range. \label{fig-1}}
\end{figure}

\begin{figure}
\epsscale{0.6} 
\plotone{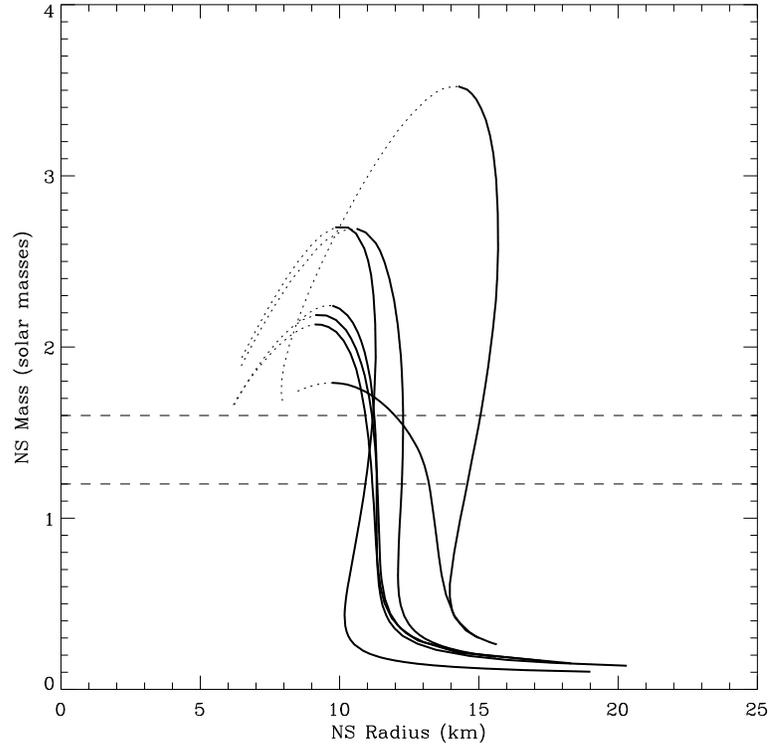} \figcaption[f2.eps]{Mass vs.~radius for NS models
constructed from seven high-density EOSs. Stable branches are
indicated with bold lines.  The dashed horizontal lines correspond to
the mass range $1.4 \pm 0.2~M_{\odot}$, which is assumed to be a
realistic NS mass range. \label{fig-2}}
\end{figure}

Figure \ref{fig-1} shows NS mass $M \equiv m(R)$ as a function of
central density $\rho_{c} \equiv \rho(0)$ for our models constructed
using the seven different EOSs described above. NS branches with $dM /
d\rho_{c} > 0$ satisfy a necessary stability criterion (e.g., Arnett
\& Bowers 1977\nocite{ab77}) and are indicated with solid bold
lines. For all EOSs used, there are stable model stars in the mass
range $1.4 \pm 0.2~M_{\odot}$. Figure \ref{fig-2} shows NS mass $M$ as
function of radius $R$ for the seven EOSs used.  Stable NS branches
are again indicated with solid bold lines. None of the models in the
stable branches violates the required stability condition $M/R < 4
c^{2}/ 9G$ for static relativistic stars \citep{g00}.

Figure \ref{fig-3} shows the ratio of the moment of inertia of the
core, $I_{\rm core}$, to the total moment of inertia, $I_{\rm total}$,
as a function of NS mass $M$.  For stable models in the mass range of
interest ($1.4 \pm 0.2~M_{\odot}$), this ratio is confined to $I_{\rm
core} / I_{\rm total} \approxgt 0.55$.  Starquakes cannot be
responsible for producing glitches in a pulsar with measured values of
$Q$ consistently outside this range unless the pulsar mass is
significantly smaller than expected.

Glitch measurements (including measurements of $Q$) for the Crab and
Vela pulsars have been published in various places in the literature
(see Table \ref{tbl-1} for references). However, no comprehensive
current listing of all measured values of $Q$ for Crab and Vela
glitches exists. Table \ref{tbl-1} is a compilation of all measured
values of $Q$ published to date for glitches from the Crab and Vela
pulsars. Some are different measurements of the same glitch. This
complete set of measurements can be used to derive a range of observed
$Q$ for each pulsar. From the 21 measurements of $Q$ for Crab
glitches, a weighted mean of the values yields $Q = 0.72 \pm
0.05$\footnote{For measurements of $Q$ without a quoted uncertainty,
an uncertainty of 0.1 was simply assumed in the calculation of the
weighted mean.}. This is only slightly smaller than the unweighted
mean of $Q = 0.83$. A range of $Q \approxgt 0.7$ encompasses the
observed distribution for the Crab pulsar.  A weighted mean of the 11
measurements of $Q$ for Vela glitches yields $0.12 \pm 0.07$, while an
unweighted mean gives $Q = 0.18$.  All estimates for Vela agree that
$Q$ is small, with a likely range $Q \approxlt 0.2$.

\section{Discussion}

The glitch behavior of the Crab \citep{acc+94, acc+96} and Vela
\citep{acc+93, cmn+93} pulsars has been extensively studied
previously.  Below we discuss and compare our model results with the
observed range of $Q$ in the context of this previous work.

\subsection{Vela} 

\citet{acc+93} and \citet{cmn+93} have found that their model
predictions of vortex unpinning in Vela are completely consistent with
the observed Vela glitch characteristics. They deduce that the
fractional crustal moment of inertia must be greater than 2.6\% for
Vela \citep{cmn+93}. \citet{da93} have used a corresponding lower
limit of 3.4\% for Vela in an attempt to rule out soft
EOSs. \citet{lel99} employ a lower limit of 1.4\% for a similar
purpose. These estimates imply an upper limit to the fractional core
moment of inertia of $\sim 0.98$, which tends to disfavor APR(s)
models with mass $M \approxgt 1.5~M_{\odot}$, but cannot independently
constrain the Vela mass since this condition is satisfied for a wide
range of model masses (see Figure \ref{fig-3}).  The conclusions made
by \citet{acc+93} and \citet{cmn+93} that the Vela mass is probably
less than $1.4~M_{\odot}$ (and is more likely closer to
$1.2~M_{\odot}$) cannot be confirmed or ruled out.  \citet{l92}
explains one reason why starquakes cannot be the glitch mechanism for
Vela: the required oblateness is not sustainable given the large size
of Vela glitches ($\Delta \Omega / \Omega \sim 10^{-6}$). After
$\sim$~100 years, the oblateness would reach zero, and therefore
glitches could only have been sustained for $\sim$ 1\% of the current
Vela age. The time interval between Vela glitches is also inconsistent
with the much longer intervals predicted by the starquake model
\citep{acc+96} unless a solid core model is invoked for Vela (e.g.,
Canuto \& Chitre 1973\nocite{cc73}; Pines, Shaham, \& Ruderman
1974\nocite{psr74}). Starquakes would also produce a thermal energy
dissipation during the large Vela glitches which is expected to be
observable as a change in X-ray luminosity soon after the glitch
occurs. X-ray observations of Vela show no such signal down to limits
of less than a few percent in the fractional change in flux
\citep{saf+00, hgh01}. The vortex unpinning model does not have this
energy dissipation problem for Vela \citep{a95}.  The low values of
$Q$ measured for Vela glitches are additional evidence that the
predictions of the starquake model do not match observations;
otherwise the implied Vela mass would be too low ($M \approxlt
0.5~M_{\odot}$ for $Q \approxlt 0.2$ in our models).  Our model
results confirm previous conclusions that starquakes cannot be a
feasible glitch mechanism for the Vela pulsar.

\subsection{Crab}

\citet{acc+94, acc+96} have studied the behavior of Crab glitches and
have determined that the starquake model is at least partially
responsible for them. The magnitudes of Crab glitches, combined with
the observed glitch rate, support this notion. \citet{l92} indicates
that the small size of Crab glitches ($\Delta \Omega / \Omega \sim
10^{-8}$) allows for small changes in oblateness which would not
significantly deplete the oblateness over the current lifetime of the
Crab. \citet{lfe98} and \citet{fle00} also show that the permanent
post-glitch offsets in the period derivative seen for the Crab pulsar
can be accounted for by starquakes: the net torque on the star
increases through shearing effects. Our model results are also
consistent with the starquake interpretation for Crab glitches. The
fractional moment of inertia values in our models are comparable to
observed $Q$ values for Crab glitches ($Q \approxgt 0.7$) if $M
\approxgt 0.15~M_{\odot}$, which seems likely. \citet{acc+94} set a
lower limit of 0.2\% for the fractional crustal moment of inertia.
This is not constraining in our models since the corresponding fractional
core moment of inertia range $I_{\rm core} / I_{\rm total} \approxlt
0.998$ is easily satisfied. Our results support previous suggestions
that starquakes could be responsible for Crab glitches.

\section{Conclusions}

\begin{figure}
\epsscale{0.6} 
\plotone{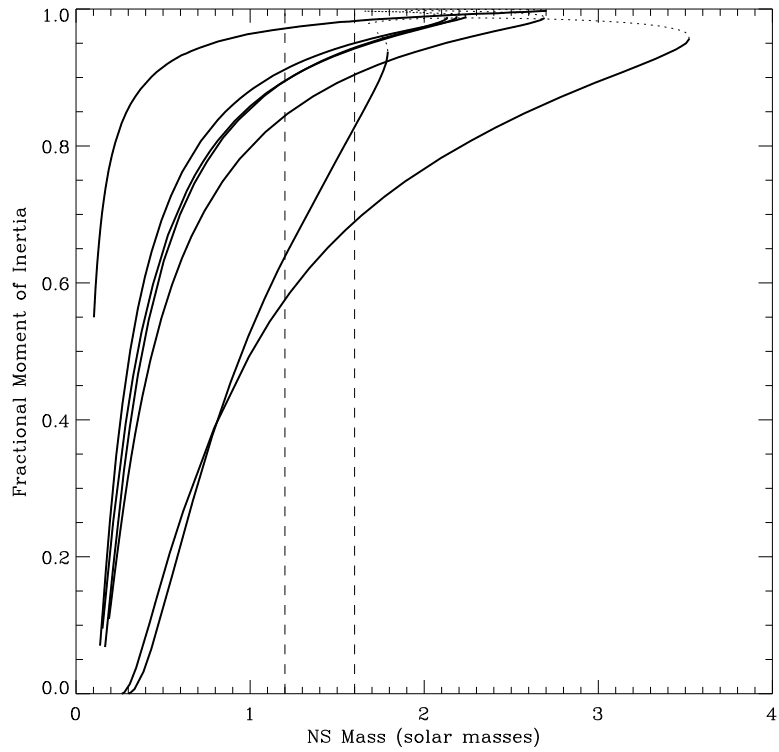} \figcaption[f3.eps]{Fractional core moment of inertia
vs.~mass for NS models constructed from seven high-density EOSs.  The
assumed realistic NS mass range $1.4 \pm 0.2~M_{\odot}$ is indicated
by the dashed vertical lines.  Stable NS configurations (in bold) in
the mass range $1.4 \pm 0.2~M_{\odot}$ have corresponding $I_{\rm
core}/I_{\rm total} \approxgt 0.55$ in all cases. This ratio is
consistent with the large values of the glitch healing parameter $Q$
predicted by the starquake model for Crab glitches ($Q \approxgt
0.7$), supporting the starquake glitch interpretation.  The ratio is
inconsistent with the much lower values of $Q$ seen for Vela ($Q
\approxlt 0.2$), indicating that starquakes do not account for Vela
glitches. \label{fig-3}}
\end{figure}

Using parameterized NS models produced from seven representative EOSs
of superdense matter, we find that the fractional moment of inertia of
the core component of the model stars is $I_{\rm core} / I_{\rm total}
\approxgt 0.55$ for all stable configurations in the assumed realistic
NS mass range $1.4 \pm 0.2~M_{\odot}$ (Figure \ref{fig-3}). This
ratio, which is predicted by the starquake model to equal the glitch
healing parameter $Q$, is not consistent with the observed range $Q
\approxlt 0.2$ for Vela glitches (see Table \ref{tbl-1}), unless the
Vela pulsar mass is unrealistically small ($M \approxlt
0.5~M_{\odot}$). This confirms results from previous studies of the
Vela pulsar which indicate that starquakes are not the cause of Vela
glitches \citep{l92, acc+93, cmn+93, a95}. The much larger values of
$Q \approxgt 0.7$ seen for Crab glitches (Table \ref{tbl-1}) are
consistent with the moment of inertia values of our models for
realistic masses, as predicted by the starquake model. These results
support previous conclusions from the analysis of Crab glitch behavior
\citep{acc+94, acc+96, lfe98, fle00} in which starquakes have been
proposed as the Crab glitch mechanism. Repeated and accurate
measurements of $Q$ for other glitching pulsars in the future (if
obtainable) could be used as a simple test of the starquake glitch
model and may help resolve whether differences between Crab-like and
Vela-like glitches can be understood on evolutionary grounds (e.g.,
Alpar 1995\nocite{a95}).

\acknowledgments

We thank the referee for helpful comments to improve the manuscript
and for suggesting the inclusion of additional equations of state in
the modeling. MD was partially supported by a grant of the Polish Committee 
for Scientific Research.

\begin{deluxetable}{lcll}
\footnotesize
\tablecaption{Measured Values of $Q$ for Crab and Vela Glitches. \label{tbl-1}}
\tablehead{
\colhead{Pulsar} &
\colhead{MJD of Glitch} & 
\colhead{$Q$} &
\colhead{Reference} 
}
\startdata
Crab           & 40493   & 0.923 $\pm$ 0.073 & Boynton et al. (1972)\nocite{bgh+72}                       \\   
(PSR B0531+21) &         & 0.93  $\pm$ 0.05  & L\"{o}hsen (1975)\nocite{l75}                              \\  
               &         & 0.94  $\pm$ 0.01  & L\"ohsen (1981)\nocite{l81}                                \\
               &         & 0.58              & Lyne, Pritchard, \& Smith (1993)\nocite{lps93}             \\ 
               & 41163   & 0.92  $\pm$ 0.02  & L\"ohsen (1981)\nocite{l81}                                \\
               & 41250   & 0.96 (fixed)      & L\"ohsen (1975)\nocite{l75}                                \\  
               &         & 0.71  $\pm$ 0.02  & L\"ohsen (1981)\nocite{l81}                                \\
               & 42447   & 0.96  $\pm$ 0.03  & L\"ohsen (1975)\nocite{l75}                                \\  
               &         & 0.77              & Lyne, Pritchard, \& Smith (1993)\nocite{lps93}             \\ 
               &         & 0.707 $\pm$ 0.002 & L\"ohsen (1981)\nocite{l81}                                \\
               & 43023   & 0.7 (fixed)       & L\"ohsen (1981)\nocite{l81}                                \\
               & 43768   & 0.7 (fixed)       & L\"ohsen (1981)\nocite{l81}                                \\
               & 44900   & $--$              & Lyne, Pritchard, \& Smith (1993)\nocite{lps93}             \\  
               & 46664   & 1.00              & Lyne, Pritchard, \& Smith (1993)\nocite{lps93}             \\ 
               & 47767   & 0.89              & Lyne, Pritchard, \& Smith (1993)\nocite{lps93}             \\ 
               & 48947   & 0.87              & Wong, Backer, \& Lyne (2001)\nocite{wbl01}      \\
               & 50021   & 0.80              & Wong, Backer, \& Lyne (2001)\nocite{wbl01}      \\
               & 50260   & 0.68              & Wong, Backer, \& Lyne (2001)\nocite{wbl01}      \\
               & 50459   & 0.87              & Wong, Backer, \& Lyne (2001)\nocite{wbl01}      \\
               & 50489   & $--$              & Wong, Backer, \& Lyne (2001)\nocite{wbl01}      \\
               & 50813   & 0.92              & Wong, Backer, \& Lyne (2001)\nocite{wbl01}      \\
               & 51452   & 0.83              & Wong, Backer, \& Lyne (2001)\nocite{wbl01}      \\
               & 51741   & 0.80 $\pm$ 0.4    & Wang et al. (2001)\nocite{wwm+01}       \\
               &         &                   &                                         \\ 
\hline
                 &       &                   &                                         \\ 
Vela             & 40280 & 0.034 $\pm$ 0.01  & Downs (1981)\nocite{d81}                \\  
(PSR B0833$-$45) & 41192 & 0.035 $\pm$ 0.001 & Downs (1981)\nocite{d81}                \\  
                 & 41308 & 0.55  $\pm$ 0.21  & Downs (1981)\nocite{d81}                \\  
                 & 42683 & 0.088 $\pm$ 0.008 & Downs (1981)\nocite{d81}                \\  
                 &       & 0.323 $\pm$ 0.012 & Manchester et al. (1983)\nocite{mng+83} \\  
                 & 43681 & 0.024 $\pm$ 0.005 & Downs (1981)\nocite{d81}                \\  
                 &       & 0.220 $\pm$ 0.036 & Manchester et al. (1983)\nocite{mng+83} \\  
                 & 44888 & 0.177 $\pm$ 0.001 & McCulloch et al. (1983)\nocite{mhr+83}  \\   
                 & 45192 & 0.044 $\pm$ 0.003 & McCulloch et al. (1987)\nocite{mkh+87}  \\
                 & 46257 & 0.158 $\pm$ 0.001 & McCulloch et al. (1987)\nocite{mkh+87}  \\
                 & 47520 & $--$              & Lyne, Shemar, \& Smith (2000)\nocite{lss00} \\
                 & 48457 & $--$              & Lyne, Shemar, \& Smith (2000)\nocite{lss00} \\
                 & 48550 & $--$              & Wang et al. (2000)\nocite{wmp+00}           \\ 
                 & 49559 & $--$              & Lyne, Shemar, \& Smith (2000)\nocite{lss00} \\
                 & 49591 & $--$              & Lyne, Shemar, \& Smith (2000)\nocite{lss00} \\
                 & 50369 & 0.38 $\pm$ 0.02   & Wang et al. (2000)\nocite{wmp+00}           \\ 
                 & 51559 & $--$              & Dodson, McCulloch, \& Lewis (2002)\nocite{dml02} \\
\enddata

\end{deluxetable}

\end{document}